\documentclass[12pt,tightenlines]{article}
\usepackage{amsmath, amssymb, amsthm,latexsym, mathrsfs}
\usepackage[usenames,dvipsnames]{color}
\usepackage[initials, alphabetic]{amsrefs}
\newcommand{\rd}{\mbox{$\rm d$}}
\newcommand{\PR}{\mathbb{P}}

\newcommand{\E}{\mathbb{E}}
\newcommand{\B}{\mathbb{B}}
\newcommand{\F}{\mathcal{F}}

\newcommand{\R}{\mathbb{R}}

\newcommand{\nn}{\nonumber}
\newtheorem{proposition}{Proposition}

\theoremstyle{definition}
\newtheorem{definition}{Definition}
\newtheorem{remark}{Remark}

\numberwithin{equation}{section}

\title{\bf{Heat Kernel Interest Rate Models with Time-Inhomogeneous Markov Processes} }
\begin{document}

\author{Jir\^o Akahori$^{\ast}$ \& Andrea Macrina$^{\dag\,\ddag}$
}
\date{}
\maketitle
\begin{center}
$^{\ast}$ Department of Mathematical Sciences, Ritsumeikan University\\
$^{\dag}$ Department of Mathematics, King's College London\\
$^{\ddag}$ Institute of Economic Research, Kyoto University
\end{center}

\begin{abstract}
%\begin{center}
We consider a heat kernel approach for the development of stochastic pricing kernels. The kernels are constructed by positive propagators, which are driven by time-inhomogeneous Markov processes. We multiply such a propagator with a positive, time-dependent and decreasing weight function, and integrate the product over time. The result is a so-called weighted heat kernel that by construction is a supermartingale with respect to the filtration generated by the time-inhomogeneous Markov processes. As an application, we show how this framework naturally fits the information-based asset pricing framework where time-inhomogeneous Markov processes are utilized to model partial information about random economic factors. We present examples of pricing kernel models which lead to analytical formulae for bond prices along with explicit expressions for the associated interest rate and market price of risk. Furthermore, we also address the pricing of fixed-income derivatives within this framework.
%\end{center}
\end{abstract}
\vspace{.25cm}
\vspace{.25cm}
\begin{center}
This version: 12 November 2010.
\\ \vspace{.25cm}
Email: akahori@se.ritsumei.ac.jp, andrea.macrina@kcl.ac.uk
\end{center}

\section{Introduction}
In work by Hughston \& Macrina (2009) and Macrina \& Parbhoo (2010), the information-based asset pricing framework developed by Brody, Hughston \& Macrina (2007, 2008a) is applied to the modelling of the pricing kernel with the goal to price fixed-income securities. The view taken is that interest rates and market prices of risk fluctuate due to noisy information about economic factors that becomes available to investors as time passes. Another reason for proposing information-based pricing kernels is the necessity to develop stochastic discount models which naturally combine with, e. g., the equity, credit and insurance pricing models presented in Brody, Hughston \& Macrina (2007, 2008a, 2008b) and in Hoyle, Hughston \& Macrina (2010). A first, albeit different, extension of information-based pricing to include stochastic discount factors can be found in Rutkowski \& Yu (2007). The theory of pricing kernels is treated in, e. g., Cochrane (2005), and in Duffie (2001).

We briefly summarise the information-based method for the construction of pricing kernels. The financial market is modelled by a filtered probability space $(\Omega, \F, \{\F_t\}, \PR)$ where $\PR$ is the real probability measure. The filtration $\{\F_t\}_{0\le t}$ is assumed to be generated by so-called information processes which model the noisy information about economic factors that market investors may observe. An information process $\{L_{tU}\}_{0\le t\le U}$ is constructed such that the associated economic factor $X_U$ is revealed at some fixed future time $U$. In order to ensure that the derived asset price processes are adapted to the market filtration, it is postulated that the pricing kernel $\{\pi_t\}$ be a functional of the information processes which generate $\{\F_t\}$. 

We shall consider the simpler situation where the filtration is generated by a single information process that has the Markov property, and where the pricing kernel can be expressed by a positive function of time and the value of the information process at that time:
\begin{equation}
 \pi_t=\pi(t,L_{tU}).
\end{equation}
The function $\pi(t,L_{tU})$ needs to be specified in such a way that the resulting pricing kernel process is a positive supermartingale. To find explicit pricing kernel models, it is also necessary to specify the information process. For example one can choose $\{L_{tU}\}$ to be a Brownian bridge information process defined by
\begin{equation}\label{BB info processes}
 L_{tU}=\sigma\,t\,X_U+\beta_{tU}
\end{equation}
where $\sigma$ is a constant parameter and $\{\beta_{tU}\}_{0\le t\le U}$ is an independent standard Brownian bridge. In this case it is convenient to assume that the pricing kernel is expressed by 
\begin{equation}\label{PK-BBform}
 \pi_t=\pi(t,L_{tU})=M_t\,f(t,L_{tU}),
\end{equation}
where $\{M_t\}_{0\le t<U}$ is a positive $\PR$-martingale that induces a change of measure to the so-called bridge measure $\B$ under which $\{L_{tU}\}$ has the law of a Brownian bridge. It is then shown in Hughston \& Macrina (2009) that $\{\pi_t\}$ is a positive $\PR$-supermartingale if and only if the function $f(t,x)$ is positive and it satisfies the following partial differential inequality:
\begin{equation}\label{ineq}
\frac{x}{U-t}\,\partial_x\,f(t,x)-
\frac{1}{2}\,\partial_{xx}\,f(t,x)-\partial_t\,f(t,x)>0.
\end{equation}
The short rate of interest $\{r_t\}$ and the market price of risk $\{\lambda_t\}$ associated with a pricing kernel of the form (\ref{PK-BBform}) are given by
\begin{eqnarray}\label{BBshortrate}
 r_t&=&\frac{1}{f(t,x)}\left(\frac{x}{U-t}\,\partial_x\,f(t,x)-\frac{1}{2}\,\partial_{xx}\,f(t,x)-\partial_t\,f(t,x)\right)\bigg\vert_{x=L_{tU}},\\
\nn\\ \label{BBmpr}
 \lambda_t&=&\frac{\sigma\,U}{U-t}\,\E\left[X_U\,\vert\,L_{tU}\right]-\frac{\partial_x\,f(t,x)}{f(t,x)}\bigg\vert_{x=L_{tU}}.
\end{eqnarray}
The condition (\ref{ineq}) is equivalent to requiring that the short rate of interest be a positive process. This is evident if (\ref{BBshortrate}) is compared with (\ref{ineq}). The expectation in (\ref{BBmpr}) can be worked out analytically by utilizing Bayes formula. 

In this paper we present a method that enables us to produce pricing kernel models which by construction are supermartingales adapted to the market filtration generated by information processes. We then show, as an example, that in the case of pricing kernels driven by Brownian bridge information processes (\ref{BB info processes}), the function $f(t,x)$ associated with the constructed pricing kernel automatically satisfies the partial differential inequality (\ref{ineq}). Thus we also obtain a class of positive interest rate models which are driven by time-inhomogenous Markov processes. The heat kernel approach for the construction of supermartingales that we consider, is introduced in Akahori, Hishida, Teichmann \& Tsuchiya (2009) for the case that the pricing kernel is driven by time-homogeneous Markov processes. We point to references therein for further background on the modelling of pricing kernels, and highlight work by Rogers (1997) and material in Hunt, Kennedy \& Pelsser (2000). Since we generate the market filtration by information processes which by construction are time-inhomogeneous Markov processes, we need to modify the heat kernel approach used by Akahori {\it et al} (2009). We focus on the weighted heat kernel approach that for our purpose turns out to be perhaps the more natural of the heat kernel methods. The Brownian bridge information process, the gamma bridge information process introduced in Brody, Hughston \& Macrina (2008b), and more generally the L\'evy random bridge information processes proposed in Hoyle, Hughston \& Macrina (2010), are all examples of time-inhomogeneous processes that possess the Markov property with respect to their natural filtration. 

We shall denote a general time-inhomogeneous Markov process by $\{X_t\}$. A L\'evy random bridge is denoted by $\{L_{tU}\}_{0\le t\le U}$ where the subscript reminds us that $\{L_{tU}\}$ has a stochastic dynamics for $t\in[0,U]$, and that after time $U$ it ``freezes'' at the value $L_{UU}$ for all $t>U$. For example, the Brownian bridge information process (\ref{BB info processes}) takes the value $L_{UU}=\sigma\,U\,X_U$ for all $t\ge U$. In Section 2 we present the heat kernel approach adapted for time-inhomogeneous Markov processes and introduce weighted heat kernels which will form the basis for the construction of information-based pricing kernels. In Section 3 we summarise the theory of L\'evy random bridges and apply it to the modelling of pricing kernels by use of weighted heat kernels. Furthermore we provide the dynamics of bond prices, and show how to price fixed income vanilla options. In Section 4 we focus on the situation where the time-inhomogeneous Markov process is a Brownian bridge information process. This example is developed in detail, offering analytical pricing kernel models. We also provide the associated explicit stochastic interest rate and market price of risk models. 
\section{Heat kernels for supermartingales}
The heat kernel approach introduced in Akahori {\it et al.} (2009)  is a systematic method for the construction of Markov functionals for interest rate models. Pricing kernel processes are modelled by assuming that they are functionals of Markov processes which generate the market filtration. The pricing kernel models give rise to the Markov functionals of the related interest rate models. Bond pricing and in general the pricing of fixed-income instruments is performed by applying the pricing kernel models. We propose the same scheme for the construction of pricing kernels driven by information processes. However, we adapted the heat kernel scheme to the case in which the driving Markov processes are time-inhomogeneous. 

Let $\{X_t\}_{0\le t}$ be a Markov process, and let $q(s,t,x,y)$ be its transition function, where $0\le s\le t$. Then we have
\begin{equation}
\PR[X_{t+s}\in \rd y\,\vert\,X_s=x]=q(s,t,x,\rd y).
\end{equation}
A Markov process is said to be time-homogeneous (or simply homogeneous)
if its transition function is independent of $s$ (see for instance Revuz \& Yor (1999)). 
If the transition function admits the form
\begin{equation}
 q (s,t,x,\rd y) = p (s,t,x,y)\rd y,  
\end{equation}
then the function $p(s,t,x,y)$ is often called a ``heat kernel''. Notice that $p(s,t,x,y)$ is usually not defined on the set $ s=t $. For the construction of pricing kernels, we consider so-called ``propagators'' defined as follows:
\begin{definition}
Let $\mathcal{S}$ be a Polish space, and $\mathcal{U}:=\{(u,t)\in[0,U)^2:u+t<U,u\neq 0\}$. 
Let $\{X_t\}_{0\le t<U}$ be a Markov process with state space $\mathcal{S}$. A measurable function $p:\mathcal{U}\times\mathcal{S}\to\R$ is a propagator if it satisfies
\begin{equation}\label{propagation}
\E[p(u,t,X_t)\,\vert\,X_s=x]=p(u+t-s,s,x)
\end{equation}
for arbitrary $(u,t)\in\mathcal{U}$ and $0\leq s\leq t$.
\end{definition}
\begin{remark}
By excluding $u=0$ from $\mathcal{U}$, we allow for the class of propagators to include heat kernels. 
\end{remark}
\begin{proposition}\label{propagator1}
Let $F:[0,U)\to\R^+$ be a measurable function. Set 
\begin{equation}\label{semigroup1}
p(u,t,x):=\E[F(u+t,X_{u+t})\,\vert\,X_t=x], 
\end{equation}
where $x\in\mathcal{S}$ and $0\le t, u$. Then $p(u,t,x)$ is a propagator. 
\end{proposition}

\noindent {\bf Proof}. We have:
\begin{eqnarray}
\E[p(u,t,X_t)\,\vert\,X_s]&=&\E\left[\E[F(u+t,X_{u+t})\,\vert\,X_t]\,\vert\,X_s\right],\nn\\
			  &=&\E[F(u+t, X_{u+t})\,\vert\,X_s],\nn\\
			  &=&\E[F((u+t-s)+s,X_{(u+t-s)+s})\,\vert\,X_s],\nn\\
			  &=&p(u+t-s,s,X_s).
\end{eqnarray}
\qed
\begin{remark}
If we allow $u=0$ in (\ref{propagation}), then the converse of 
Proposition \ref{propagator1} holds. 
Namely, the propagator is represented by some measurable function $F:[0,U)\to\R^+$ as given in (\ref{semigroup1}). More precisely, we have
\begin{equation}
p(u,t,X_t)=\E[p(0,u+t,X_{u+t})\,\vert\,X_t],
\end{equation}
by letting $u\to 0$, $t \to u+t$ and $s \to t$. That is, $F(t,x)=p(0,t,x)$. 
\end{remark}
\begin{definition}
Let a measurable function $w:[0,U]^2 \rightarrow \R^+$ that satisfies 
\begin{equation}\label{weight}
w(t,u-s)\leq w(t-s,u)
\end{equation}
for $0<U\le\infty$ and $s\leq t\wedge u$, be called a ``weight function''. 
A ``weighted heat kernel'' $f(t,x)$ is defined by
\begin{equation}
 f(t,x)=\int_{0}^{U-t}p(u,t,x)\,w(t,u)\,\rd u,
\end{equation}
for $0\le t<U$, $u\ne 0$ and where $p(u,t,x)$ is a propagator. 
\end{definition}
\begin{proposition}\label{supermart prop}
Consider the propagator (\ref{semigroup1}). 
Assume that $\PR[f(t,X_t)<\infty]=1$ for all $t\in[0,U)$, and let $0\le u+t< U$. Then 
\begin{equation}\label{supermart}
f(t,X_t)=\int_0^{U-t}\E\left[F(u+t,X_{u+t})\,\vert\,X_t\right]\,w(t,u)\,\rd u
\end{equation}
is a positive supermartingale.
\end{proposition}
\begin{proof}
For $0\le s\le t$, we have
\begin{equation}\label{cond-supmart}
\E[f(t,X_t)\,\vert\,X_s]=\int_0^{U-t}\E[\E\left[F(u+t,X_{u+t})\,\vert\,X_t\right]\,\vert\,X_s]\,w(t,u)\rd u 
\end{equation}
where we make use of Fubini's theorem. By the propagation property and the change of variables $u=v-t+s$, we obtain
\begin{equation}\label{pkcondexp} 
\E[f(t,X_t)\,\vert\,X_s]=\int_{t-s}^{U-s}\E\left[F(v+s,X_{v+s})\,\vert\,X_s\right]\,w(t,v-t+s)\rd v.
\end{equation}
We complete the proof by applying property (\ref{weight}):
\begin{eqnarray}
\E[f(t,X_t)\,\vert\,X_s]
&\leq&\int_{t-s}^{U-s}\E\left[F(v+s,X_{v+s})\,\vert\,X_s\right]\,w(t-(t-s),v)\rd v\nn\\
&\leq&\int_0^{U-s}\E\left[F(v+s,X_{v+s})\,\vert\,X_s\right]\,w(s,v)\rd v\nn\\
\nn\\
&\leq&f(s,X_s).
\end{eqnarray}
\end{proof}
\begin{remark}\label{pweight}
(i) Let $w_1(t,u)$ and $w_2(t,u)$ be weight functions, and $c$ be a positive constant. 
Then $c\,w_1(t,u)$, $w_1(t,u)+w_2(t,u) $, and $w_1(t,u)\,w_2(t,u)$ are weight functions.   
(ii) If $w(t,u)$ is non-increasing in $t$ and non-decreasing in $u$,
then $w(t,u)$ is a weight function.
(iii) If $w(t,u)=\bar{w}(t+u)$ for some measurable function $\bar{w}:\R\to\R^+$,
then $w(t,u)$ is a weight function. 

The assertions (i) and (ii) are obvious. The assertion (iii) is shown by 
$w(t,u-s)=\bar{w}(t+u-s)=w(t-s, u)$. 
\end{remark}
\begin{remark}
Since the state space of the time-inhomogeneous Markov processes is a Polish space, the heat kernel approach may also be applied to construct pricing kernel models which are driven by multiple information processes. The development of explicit multi-factor models is deferred to future research. We refer to Hughston \& Macrina (2009) and Macrina \& Parbhoo (2010) for further material on multi-factor interest rate models,  and emphasize that multi-factor pricing kernels may have dynamics which is simultaneously driven by time-inhomogeneous Markov processes based, for instance, on various L\'evy processes. 
\end{remark}
\section{Pricing with time-inhomogeneous Markov \\ information}\label{Pricing with TIH MI}
Fixed-income assets can be priced by use of pricing kernel models. Let us consider a single-dividend paying asset which, at a fixed future date $T$, has a random cash flow $H_T$. The price $H_t$ at time $t\le T$ of such an asset is
\begin{equation}\label{pricing formula}
H_t=\frac{1}{\pi_t}\,\E\left[\pi_T\,H_T\,\vert\,\F_t\right],
\end{equation}
where $\{\pi_t\}$ is the pricing kernel. For $H_T=1$ we recover the pricing formula for a discount bond with maturity $T$. In order to obtain specific processes for the discount bond system, we need to develop explicit pricing kernel models. Denote the price process of the discount bond by $\{P_{tT}\}_{0\le t\le T}$, and consider pricing kernel models of the form $\pi_t=\pi(t,X_t)$ where $\{X_t\}$ is a Markov process that generates the market filtration $\{\F_t\}$. Hence we have
\begin{equation}\label{bond price model}
 P_{tT}=\frac{\E\left[\pi(T,X_T)\,\vert\,X_t\right]}{\pi(t,X_t)}.
\end{equation}
No-arbitrage requires the pricing kernel to be a positive supermartingale. We can thus apply Proposition \ref{supermart prop} to construct these pricing kernels:  
\begin{equation}\label{gensupmartin}
 \pi(t,X_t)=\int_0^{U-t}\E\left[F(u+t,X_{u+t})\,\vert\,X_t\right]\,w(t,u)\,\rd u.
\end{equation}
Assuming $0\le t\le T<U$, the price of the discount bond is
\begin{equation}\label{discbondprice}
 P_{tT}=\frac{\int^{U-t}_{T-t}\E\left[F(u+t,X_{u+t})\,\vert\,X_t\right]\,w(T,u-T-t)\,\rd u}{\int^{U-t}_0\E\left[F(u+t,X_{u+t})\,\vert\,X_t\right]\,w(t,u)\,\rd u},
\end{equation}
where (\ref{pkcondexp}) is used in order to work out the conditional expectation in (\ref{bond price model}).

There are three ingredients which need to be specified so that explicit pricing models can be derived: 1) the Markov process $\{X_t\}$, 2) the positive function $F(t,x)$, and 3) the weight function $w(t,u)$. The specification of $\{X_t\}$ is rather crucial since it also determines the model for the market filtration. As an example of a class of Markov processes which can be applied to model the market information flow, we take the L\'evy random bridges proposed in Hoyle, Hughston \& Macrina (2010). These Markov processes are manifestly time-inhomogeneous since their transition function depends explicitly on the time variable. We shall adopt the notation in Hoyle {\it et al}. (2010), that is also useful in making a distinction between a general Markov process $\{X_t\}$ and a L\'evy random bridge denoted $\{L_{tU}\}$.
\begin{definition}
 A L\'evy random bridge $\{L_{tU}\}_{0\le t\le U}$ that takes values in a continuous state space, is a process with the following properties:
\begin{enumerate}
 \item $L_{UU}$ has marginal law $\nu$.
 \item There exists a L\'evy process $\{L_t\}$ such that $L_t$ has density $\rho_t(x)$ for all $t\in(0,U]$.
 \item  $\nu$ concentrates mass where $\rho_U(z)$ is positive and finite, i.e.~$0<\rho_U(z)<\infty$ for $\nu$-almost every~$z$.
 \item For every $n\in\mathbb{N}_+$, every $0<t_1<\cdots<t_n<U$, every $(x_1,\ldots,x_n)\in\R^n$, and $\nu$-almost every~$z$, we have
    \begin{align*}
      &\PR\left[L_{t_1,U}\leq x_1,\ldots,L_{t_n,U} \leq x_n \left\vert\, L_{UU} = z \right.\right]\\
        &\hspace{4.5cm}=\PR\left[L_{t_1}\leq x_1,\ldots,L_{t_n} \leq x_n \left\vert\, L_{U} = z \right.\right].
    \end{align*}
\end{enumerate}
\end{definition}
For the computation of the conditional expectation involved in (\ref{gensupmartin}) and (\ref{discbondprice}), the following result in Hoyle {\it et al}. (2010) is useful: For $0\le s<t<U$,
\begin{equation}\label{L-ConditionalDens}
\PR\left[L_{tU}\in\rd
y\,\vert\,L_{sU}=x\right]=\frac{\psi_{tU}(\R;y)}{\psi_{sU}(\R,x)}\,\rho_{t-s}(y-x)\,\rd y,
\end{equation}     
where  
\begin{align}
&\psi_{tU}(\R;y):=\int^{\infty}_{-\infty}\frac{\rho_{U-t}(z-y)}{\rho_U(z)}\,\nu(\rd
z),& &\psi_{0U}(\rd z;y):=\nu(\rd z).&
\end{align}
If we assume that the Markov process $\{X_t\}$ driving the bond price process (\ref{discbondprice}) is a L\'evy random bridge, then the conditional expectation in (\ref{discbondprice}) is given by
\begin{equation}\label{L-propagator}
\E\left[F(u+t,L_{u+t,U})\,\vert\,L_{tU}\right]=\int^{\infty}_{-\infty}F(u+t,y)\frac{\psi_{u+t,U}(\R;y)}{\psi_{tU}(\R,L_{tU})}\,\rho_{u}(y-L_{tU})\,\rd y.
\end{equation}
In this case, the evolution of the bond price is determined by the conditional density $\rho_t$ of the L\'evy process $\{L_t\}$. The density $\rho_t$ is given by a function of time $t$ and the value of $L_{tU}$ prevailing at $t$.   

Before we go on to work out explicitly the price of a fixed-income security, e.g. a discount bond, by choosing a particular L\'evy random bridge, we show how in this framework one can also price derivatives. In particular, we show how to calculate a bond option via (\ref{pricing formula}). Let $\{C_{st}\}$, $0\le s\le t\le T<U$, be the
price process of a European-style call bond option with maturity $t$ and strike $K$. 
The price at time $s$ is
\begin{equation}\label{optionproc}
C_{st}=\frac{1}{\pi(s,L_{sU})}\,
\E^{\PR}\left[\pi(t,L_{tU})\left(P_{tT}-K\right)^+\,\vert\,L_{sU}\right],
\end{equation}
where here $P_{tT}$ is defined by (\ref{bond price model}) for $X_t=L_{tU}$. Since the pricing kernel is a positive process, we may take $\pi(t,L_{tU})$ inside the $\max$-function. We obtain  
\begin{equation}
C_{st}=\frac{1}{\pi(s,L_{sU})}\,\E^{\PR}\left[\left(\E^{\PR}\left[\pi(T,L_{TU})\,\vert\,L_{tU}\right]-K\pi(t,L_{tU})\right)^+\,\vert\,L_{sU}\right].
\end{equation}
For $\pi(t,L_{tU})$, we use formula (\ref{gensupmartin}) where $X_t=L_{tU}$. Furthermore, 
\begin{equation}
\E^{\PR}\left[\pi(T,L_{TU})\right]=\int^{U-t}_{T-t}\E^{\PR}\left[F(u+t,L_{u+t,U})\,\vert\,L_{tU}\right]w(T,u-T-t)\rd u,
\end{equation}
which can be simplified by (\ref{L-propagator}). The implicit result is a function of time $t$ and $L_{tU}$ that we denote $I(t,L_{tU})$. We rewrite the option price in the form
\begin{align}
C_{st}=\frac{1}{\pi(t,L_{sU})}\,\E^{\PR}\left[\left(I(t,L_{tU})-K\pi(t,L_{tU})\right)^+\,\vert\,L_{sU}\right].
\end{align}
For fixed $t$, let $z^{\ast}$ be the set defined by
\begin{equation}
z^{\ast}:=\left\{z\,\vert\,I(t,z)-K\pi(t,z)>0\right\}.
\end{equation}
Then we have
\begin{equation}
C_{st}=\frac{1}{\pi(s,L_{sU})}\int^{\infty}_{z^{\ast}}\left(I(t,z)-K\,\pi(t,z)\right)\PR\left[L_{tU}\in\rd
z\,\vert\,L_{sU}\right],
\end{equation}
where $\PR\left[L_{tU}\in\rd z\,\vert\,L_{sU}\right]$ is given by (\ref{L-ConditionalDens}). We observe that the option price is determined by a function of time and the value at time $s$ of the time-inhomogeneous Markov process $\{L_{tU}\}$.
\section{Explicit pricing kernel models}
We reconsider the case in which the L\'evy random bridge is a Brownian random bridge (Brownian bridge information process) as defined in (\ref{BB info processes}). As shown in Hoyle {\it et al.} (2010), the conditional density (\ref{L-ConditionalDens}) can be computed in closed form. We can thus utilize this result in order to directly write down the expression for the bond price process (\ref{discbondprice}) driven, e. g., by Brownian bridge information. In this case we need working out a Gaussian integral over the range of the variable $u$. However, in what follows, we shall apply a change-of-measure technique introduced in Brody {et al.} (2007, 2008).  

Recall that the Brownian bridge information process is a L\'evy random bridge of the form
\begin{equation}\label{BBIP}
L_{tU}=\sigma\,t\,X_U + \beta_{tU},
\end{equation}
where $\{\beta_{tU}\}_{0\le t\le U}$ is a Brownian bridge. Let $\{M_t\}_{0\le t<U}$ be defined by
\begin{equation}\label{Mdensity}
\frac{\rd M_t}{M_t}=-\frac{\sigma
U}{U-t}\,\E\left[X_U\,\vert\,L_{tU}\right]\rd W_t,
\end{equation} 
where
\begin{equation}
W_t=L_{tU}+\int^t_0\frac{L_{sU}}{U-s}\,\rd s-\sigma
U\int^t_0\frac{1}{U-s}\,\E\left[X_U\,\vert\,L_{sU}\right]\rd s.
\end{equation}
The $(\{\F_t\},\PR)$-martingale $\{M_t\}$ induces a change of
measure from $\PR$ to the so-called bridge measure $\B$ under which
$\{L_{tU}\}$ has the distribution of a Brownian bridge. We emphasize that the process $\{W_t\}$ is an $(\{\F_t\},\PR)$-Brownian motion. The bond price is modelled by
\begin{equation}
P_{tT}=\frac{\E\left[\pi(T,L_{TU})\vert\,L_{tU}\right]}{\pi(t,L_{tU})}.
\end{equation}
Next we assume, with no
loss of generality, that
\begin{equation}\label{BB-PK}
\pi(t,L_{tU})=M_t\,f(t,L_{tU}),
\end{equation}
and perform the change of measure $\PR$ to $\B$ so that 
\begin{equation}
P_{tT}=\frac{\E^{\B}\left[f(T,L_{TU})\vert\,L_{tU}\right]}{f(t,L_{tU})}.
\end{equation}
The pricing kernel
$\{\pi(t,L_{tU})\}$ is a positive $\PR$-supermartingale if and only if $\{f(t,L_{tU})\}$
is a positive supermartingale under $\B$. That is, for $0\le s\le t$,
\begin{align}\label{supermart-proof}
\mathbb{E}^\mathbb{P}[\pi_t\,\vert\,\mathcal{F}_s] = M_s\,\mathbb{E}^{\mathbb{B}}[f(t,L_{tU})\,\vert\,L_{sU}] \leq M_s\,f(s,L_{sU})=\pi_s.
\end{align}
The positive $\B$-supermartingale $\{f(t,L_{tU})\}_{0\le t<U}$ can now be constructed by positive weighted heat kernels according to Proposition \ref{supermart prop}, where the conditional expectation is taken with respect to the $\B$-measure. The price $P_{tT}$ is thus calculated by the formula
\begin{equation}\label{BBbond}
 P_{tT}=\frac{\int^{U-t}_{T-t}\E^{\B}\left[F(u+t,L_{u+t,U})\,\vert\,L_{tU}\right]\,w(T,u-T+t)\rd u}{\int^{U-t}_0\E^{\B}\left[F(u+t,L_{u+t,U})\,\vert\,L_{tU}\right]\,w(t,u)\,\rd u}.
\end{equation}
Since $L_{tU}$ has the law of a Brownian bridge under the $\B$-measure, the conditional expectation involves the Gaussian density function. Explicit expressions for the bond price are obtained by specifying $F(t,x)$ and the weight function $w(t,u)$. Depending on the combination of $F(t,x)$ and $w(t,u)$, the integration with respect to $u$ in (\ref{BBbond}) can be performed in closed form. Some examples follow:
\\

{\bf Quadratic models}.
We consider the positive function $F(x)=x^2$, and a weight function $w(t,u)=U-t-u$. As we shall see shortly, such a combination will lead to a stochastic interest rate model and a closed-form expression for the bond price. The propagator (\ref{L-propagator}) that gives rise to the stochastic dynamics of the bond price is given by
\begin{equation}\label{quad-propagator}
 \E^{\B}\left[\left(L_{u+t,U}\right)^2\,\big\vert\,L_{tU}\right]=\frac{u(U-t-u)}{U-t}+\left(\frac{U-t-u}{U-t}\right)L_{tU}^2.
\end{equation}
Here we recall that $L_{tU}$ has the law of a Brownian bridge over $[0,U)$ under the $\B$-measure. The weighted heat kernel takes the form 
\begin{eqnarray}\label{quad-PK}
f(t,L_{tU})&=&\int^{U-t}_0\E^{\B}\left[F(L_{u+t,U})\,\vert\,L_{tU}\right]w(t,u)\,\rd u,\nn\\
&=&\frac{1}{12}\,(U-t)^3+\frac{1}{4}\,(U-t)^2\,L^2_{tU},
\end{eqnarray}
which ensures that the pricing kernel (\ref{BB-PK}) is, via (\ref{supermart-proof}), a positive $\PR$-supermartingale. Inserting (\ref{quad-propagator}) and (\ref{quad-PK}) in (\ref{BBbond}), we obtain the bond price
\begin{align}\label{quadbondprice}
P_{tT}=\frac{\frac{1}{12}\,(U-T)^3+\frac{1}{4}\,\frac{(T-t)(U-T)^3}{(U-t)}+\frac{1}{4}\,\frac{(U-T)^4}{(U-t)^2}\,L_{tU}^2}{\frac{1}{12}\,(U-t)^3+\frac{1}{4}\,(U-t)^2\,L_{tU}^2}.
\end{align}   
The associated short rate of interest $\{r_t\}$ can be obtained
by calculating the instantaneous forward rate associated with the
bond price $\{P_{tT}\}_{0\le t\le T<U}$. Alternatively, (\ref{quad-PK}) can be inserted in formula (\ref{BBshortrate}) for the short rate process. We get
\begin{equation}
r(t,L_{tU})=\frac{L_{tU}^2}{\frac{1}{4}\,(U-t)\left[\frac{1}{3}(U-t)+L_{tU}^2\right]},
\end{equation}
for $0\le t<U$. We emphasize that this is, by construction, a positive interest rate model.
The market price of risk $\{\lambda_t\}$ can be obtained via (\ref{BBmpr}). We have
\begin{equation}
\lambda(t,L_{tU})=\frac{\sigma
U}{U-t}\E^{\PR}\left[X_U\,\vert\,L_{tU}\right]-\frac{\frac{1}{2}(U-t)^2
L_{tU}}{\frac{1}{12}\,(U-t)^3+\frac{1}{4}\,(U-t)^2\,L^2_{tU}}.
\end{equation}
The expectation $\E^{\PR}\left[X_U\,\vert\,L_{tU}\right]$ can be calculated in closed form by applying Bayes formula.
\\

{\bf Exponential quadratic models}.
We consider a positive function $F(t,x)$ that depends explicitly on time, unlike in the previous example.
Let 
\begin{equation}
F(u+t,x)=\exp\left(\tfrac{1}{2}\,\gamma_{t+u}\,x^2\right), 
\end{equation}
where $\gamma_{t+u}$ is positive and deterministic. In this case the propagator takes the form
\begin{align}
&\E^{\B}\left[\exp\left(\tfrac{1}{2}\,\gamma_{t+u}\,L^2_{t+u,U}\right)\,\vert\,L_{tU}\right]\\
&\hspace{3.25cm}=\frac{1}{\sqrt{1-u\,\gamma_{t+u}\,a_{t+u}}}\,\exp\left(\frac{\gamma_{t+u}\,a^2_{t+u}}{2\left(1-u\,\gamma_{t+u}\,a_{t+u}\right)}L^2_{tU}\right),\nn
\end{align}
where $a_{t+u}=(U-t-u)/(U-t)$. By setting $ \gamma_{t+u}=(U-t-u)^{-1}, $ and by choosing a weight
function
\begin{equation}
w(t,u)=(U-t-u)^{\eta-\frac{1}{2}}\qquad(\eta>\tfrac{1}{2}),
\end{equation}
we obtain an analytical expression for the $\B$-supermartingale
$\{f(t,L_{tU})\}$:
\begin{equation}\label{expSUP}
f(t,L_{tU})=\frac{1}{\eta-\tfrac{1}{2}}\,(U-t)^{\eta}\exp\left(\frac{L^2_{tU}}{2(U-t)}\right).
\end{equation}
This supermartingale leads to a deterministic bond price, even though the related pricing kernel is stochastic.
However we can modify $\{f(t,L_{tU})\}$ slightly.
Let $g_0(t)$ and $g_1(t)$ be positive, decreasing, and differentiable
functions. Consider the supermartingale
\begin{equation}
\tilde{f}(t,L_{tU})=g_0(t)+g_1(t)(U-t)^{\eta}\exp\left(\frac{L^2_{tU}}{2(U-t)}\right).
\end{equation}
Then the associated bond price system has the following stochastic dynamics:
\begin{equation}
P_{tT}=\frac{g_0(T)+g_1(T)(U-T)^{\eta-1/2}(U-t)^{{1}/{2}}\exp\left(\frac{L^2_{tU}}{2(U-t)}\right)}
{g_0(t)+g_1(t)(U-t)^{\eta}\exp\left(\frac{L^2_{tU}}{2(U-t)}\right)},
\end{equation}
for $t\in[0,U)$ and $u\in[0,U-t]$. By use of (\ref{BBshortrate}) we work out the short rate process, and obtain 
\begin{align}\label{expquadSIR}
r_t&=\frac{g_1(t)\,(U-t)^{\eta}\,\exp\left[\frac{L_{tU}^2}{2(U-t)}\right]}{g_0(t)+g_1(t)\,(U-t)^{\eta}\,\exp\left[\frac{L_{tU}^2}{2(U-t)}\right]}\nn\\
&\times\left[\left(\eta-\tfrac{1}{2}\right)(U-t)^{-1}-\frac{\partial_t\,g_1(t)}{g_1(t)}-\frac{\partial_t\,g_0(t)}{g_1(t)\,(U-t)^{\eta}\,\exp\left[\frac{L_{tU}^2}{2(U-t)}\right]}\right].
\end{align}
Next we consider a particular choice for $g_1(t)$ . If
\begin{equation}
g_1(t)=(U-t)^{-(\eta-\frac{1}{2})},
\end{equation}
then (\ref{expquadSIR}) reduces to 
\begin{equation}\label{special expquadSIR}
r_t=-\frac{\partial_t\,g_0(t)}{g_0(t)+(U-t)^{1/2}\,\exp\left[\frac{L_{tU}^2}{2(U-t)}\right]}.
\end{equation}
Since $g_0(t)$ is a positive decreasing function, its derivative is negative. Thus the interest rate process (\ref{special expquadSIR}) is positive. The market price of risk model can of course also be derived in closed form---e. g., via (\ref{BBmpr})---and it is given in terms of a function of time and $L_{tU}$. 
We conclude this section by noting that further examples can be constructed. A semi-analytic
formula is obtained for the exponential linear family where, for instance, $F(x)=\exp(-\mu\,x)$ may be chosen.
\section{Fixed-income derivatives with Brownian \\ bridge information}\label{Sec BI-OP}
In Section \ref{Pricing with TIH MI} we introduced formula (\ref{optionproc}) for the pricing of a European-style bond option with strike $K$ and maturity $t$. The market information is modelled by a L\'evy random bridge process, and the pricing kernel is assumed to be of the form $\pi_t=\pi(t,L_{tU})$. The price process of the bond option is then given implicitly in terms of a function of time and the L\'evy random bridge.

In this section we go one step further, and price the bond option under the assumption that the market filtration is generated by a Brownian bridge information process (\ref{BBIP}). In order to derive an explicit price process for the option, we need to specify the price process of the underlying bond. We shall use the quadratic models as an example, and thus apply (\ref{quadbondprice}). The related pricing kernel model is 
\begin{equation}
\pi_t=M_t f(t,L_{tU}),
\end{equation}
where $\{M_t\}_{0\le t<U}$ is defined by (\ref{Mdensity}), and $\{f(t,L_{tU})\}$ is given by (\ref{quad-PK}). We write down equation (\ref{optionproc}) for the option price under the bridge measure $\B$, that is
\begin{equation}
C_{st}=\frac{1}{f(s,L_{sU})}
\E^{\B}\left[f(t,L_{tU})\left(P_{tT}-K\right)^+\,\vert\,L_{sU}\right].
\end{equation}
We take the positive $\B$-supermartingale $\{f(t,L_{tU})\}$ inside the $\max$-function where, on one hand it cancels with the denominator of $P_{tT}$, and on the other hand, it is multiplied with the strike $K$. We re-arrange the resulting terms, and obtain
\begin{eqnarray}
C_{st}&=&\frac{1}{f(s,L_{sU})}\E^{\B}\left[\left(\frac{1}{4}\frac{(T-t)(U-T)^3}{U-t}
+\tfrac{1}{12}\left[(U-T)^3-K(U-t)^3\right]\right.\right.\nn\\
&+&\left.\left.\frac{1}{4}\left[\frac{(U-T)^4}{(U-t)^2}-K(U-t)^2\right]\,L_{tU}^2\right)^+\,\bigg\vert\,L_{sU}\right]
\end{eqnarray}
In order to work out the conditional expectation, we make the substitution
\begin{equation}
L_{tU}=\nu_{st}\,Y+\frac{U-t}{U-s}\,L_{sU},
\end{equation}
where $Y$ is a standard normally-distributed random variable that is $\B$-independent of $L_{sU}$. Furthermore,
\begin{equation}
\nu_{st}=\sqrt{\frac{(t-s)(U-t)}{U-s}}.
\end{equation}
Due to the properties of $Y$, the expression for the option price reduces to the following Gaussian integral:
\begin{equation}
C_{st}=\frac{1}{f(s,L_{sU})}\int^{\infty}_{-\infty}\left(A+B\left(\nu_{st}\,y+\frac{U-t}{U-s}\,L_{sU}\right)^2\right)^+\frac{1}{\sqrt{2\pi}}
\,\exp\left(-\tfrac{1}{2}\,y^2\right)\rd y,
\end{equation}
where 
\begin{eqnarray}
A&=&\frac{1}{4}\frac{(T-t)(U-T)^3}{U-t}
+\tfrac{1}{12}\left[(U-T)^3-K(U-t)^3\right]\\
B&=&\frac{1}{4}\left[\frac{(U-T)^4}{(U-t)^2}-K(U-t)^2\right].
\end{eqnarray}
Next we integrate over the range of $y$ for which the $\max$-function does not vanish. 
This can be written as follows:
\begin{equation}
C_{st}=\frac{1}{f(s,L_{sU})}\int^{\infty}_{cy^2+by+a\,>\,0}\left(cy^2+by+a\right)\frac{1}{\sqrt{2\pi}}\,\exp\left(-\tfrac{1}{2}y^2\right)\rd y,
\end{equation}
where
\begin{align}
&a=A+B\left(\frac{U-t}{U-s}\right)^2 L_{sU}^2,&
&b=2B\nu_{st}\,\frac{U-t}{U-s}\,L_{sU},& &c=B\nu_{st}^2.&
\end{align}
Next we solve for the values $y$ that satisfy the
inequality $cy^2+by+a>0$. Let $N(x)$ denote the cumulative normal distribution function. Then:
\\
1) Case  $c=0$. For $b>0$ we obtain
\begin{equation}
C_{st}=\frac{1}{f(s,L_{sU})}\left[a\,N\left(\frac{a}{b}\right)+\frac{b}{\sqrt{2\pi}}\exp\left(-\tfrac{1}{2}\left(\frac{a}{b}\right)^2\right)\right],
\end{equation}
and for $b<0$ we have
\begin{equation}
C_{st}=\frac{1}{f(s,L_{sU})}\left[a\,N\left(-\frac{a}{b}\right)-\frac{b}{\sqrt{2\pi}}\exp\left(-\tfrac{1}{2}\left(\frac{a}{b}\right)^2\right)\right].
\end{equation}
\\
2) Case $c\neq0$ and $b^2-4ac = -4AB \nu_{st}^2 >0$. 
For $c<0$ we have
\begin{equation}
y_{\pm}=\frac{1}{2c}\left(-b\pm\sqrt{b^2-4ac}\right), 
\end{equation}
and thus
\begin{eqnarray}
C_{st}&=&\frac{1}{f(s,L_{sU})}\left[(a+b)\left[N(y_+)-N(y_-)\right]+(b+cy_-)
\,\exp\left(-\tfrac{1}{2}(y_-)^2\right)\right.\nn\\
&&\hspace{2cm}-\left.(b+cy_+)\,\exp\left(-\tfrac{1}{2}(y_+)^2\right)\right].
\end{eqnarray}
For $c>0$ we have either $y\le y_-$ or $y\ge y_+$ and thus
\begin{eqnarray}
C_{st}&=&\frac{1}{f(s,L_{sU})}
\left[(a+c)\left[N(y_-)+N(y_+)\right]-(b+cy_-)\frac{1}{\sqrt{2\pi}}\,
\exp\left(-\tfrac{1}{2}(y_-)^2\right)\right.\nn\\
&&\left.\hspace{2cm}+(b+cy_+)\frac{1}{\sqrt{2\pi}}\,
\exp\left(-\tfrac{1}{2}(y_+)^2\right)\right].
\end{eqnarray}
3) Case $c\neq0$ and $b^2-4ac =-4AB \nu_{st}^2 \le 0$. 
For $c<0$ the probability that
$cy^2+by+a>0$ is zero. This means that in such a case the option at
time $s$ is almost always out of the money, and we have
$C_{st}=0$. For $c>0$ we have
$cy^2+by+a>0$ with probability one, and thus the option is almost always in the money. Hence we have
\begin{equation}
C_{st}=\frac{a+c}{f(s,L_{sU})}.
\end{equation}
\newpage
\vskip 15pt \noindent {\Large \bf Acknowledgements}
\\

The authors are grateful to seminar participants at the KIER-TMU International Workshop on Financial Engineering 2009, the Vienna Institute of Finance (September 2009), Hitotsubashi University (November 2009), and at the Workshop on Financial Derivatives and Risk Management at the Fields Institute in Toronto (May 2010) for helpful comments and suggestions. In particular, we thank P. A. Parbhoo and J. Sekine for stimulating and fruitful discussions. AM thanks Ritsumeikan University in Kusatsu, Japan, for their hospitality and friendly work environment. 

\vskip 15pt \noindent {\Large \bf References}

\begin{enumerate}
 \item J.~Akahori, Y.~Hishida, J.~Teichmann \& T.~Tsuchiya (2009) A heat kernel approach to interest rate models. arXiv.org: 0910.5033

\item D.~C.~Brody, L.~P.~Hughston \& A.~Macrina (2007) Beyond hazard rates: a new framework for credit-risk modelling. In: Advances in Mathematical Finance , Festschrift volume in honour of Dilip Madan, R.~Elliot, M.~Fu, R.~Jarrow, J.~Y. Yen, editors. Birkh\"auser \& Springer Verlag.

\item D.~C.~Brody, L.~P.~Hughston \& A.~Macrina (2008a) Information-based asset pricing. International Journal of Theoretical and Applied Finance 1, 107-142.

\item D.~C.~Brody, L.~P.~Hughston \& A.~Macrina (2008b) Dam rain and cumulative gain. Proceedings of the Royal Society A 464, 1801-1822.

\item J.~H.~Cochrane (2005) \emph{Asset pricing}. Revised edition, Princeton University Press.

\item D.~Duffie (2001) \emph{Dynamic asset pricing}. Third edition, Princeton University Press. 

\item E.~Hoyle, L.~P.~Hughston \& A.~Macrina (2010a) L\'evy random bridges and the modelling of financial information. arXiv.org: 0912.3652

\item L.~P.~Hughston \& A.~Macrina (2009) Pricing fixed-income securities in an information-based framework. arXiv.org: 0911.1610

\item P.~J.~Hunt, J.~E.~Kennedy \& A.~Pelsser (2000) Markov-functional interest rate models. Finance and Stochastics {\bf 4}, No. 4, 391-408. 

\item A.~Macrina \& P.~A.~Parbhoo (2010) Security pricing with information-sensitive discounting. In: Recent Advances in Financial Engineering 2009: Proceedings of the KIER-TMU International Workshop on Financial Engineering 2009, M.~Kijima, C.~Hara, K.~Tanaka, editors. World Scientific Publishing Company. 

\item D.~Revuz \& M.~Yor (1999) \emph{Continuous martingales and Brownian motion}. 3rd edition, Springer Verlag.

\item L.~C.~G.~Rogers (1997) The potential approach to the term structure of interest rates and foreign exchange rates. Mathematical Finance {\bf 7}, No. 2, 157-176.

\item M.~Rutkowski \& N.~Yu (2007) On the Brody-Hughston-Macrina approach to modelling of defaultable term structure. International Journal of Theoretical and Applied Finance 10, 557-589.
\end{enumerate}
\end{document}